\documentclass[showpacs,amsmath,amssymb,aps]{revtex4}

\usepackage{graphicx}
\usepackage{dcolumn}
\usepackage{bm}
\usepackage{epsfig}

\begin{document}

\title{Mean-field approximation for the chiral soliton in a chiral phase transition}
\author{Hui~Zhang}
\author{Song~Shu}\email[Corresponding author.\\E-mail address:]{shus@hubu.edu.cn}
\affiliation{Department of Physics and Electronic Technology, Hubei University, Wuhan 430062, China}

\begin{abstract}
In the mean-field approximation we study the chiral soliton within the linear sigma model in a thermal vacuum. The chiral soliton equations with different boundary conditions are solved at finite temperatures and densities. The solitons are discussed before and after the chiral restoration. We find that the system has soliton solutions even after the chiral restoration and they are very different from those before the chiral restoration, which indicates that the quarks are still bounded after the chiral restoration.
\end{abstract}
\pacs{12.39.Fe, 12.38.Mh, 12.38.Aw, 11.10.Wx}
\maketitle

\section{Introduction}
There is a quantum chromodynamics (QCD) phase transition between normal nuclear matter and quark-gluon plasma (QGP) with chiral symmetry restoration and deconfinement at finite temperature and density \cite{ref1,ref2}. Theoretically, perturbative QCD calculations work well on the scale of very small distances or very high energies due to the essential properties of asymptotic freedom. However, analytical as well as numerical methods have not been developed fully for the low energy area because of the nonperturbative features of broken chiral symmetry and confinement \cite{ref3,ref4,ref5,ref6,ref7}. Both features are intimately related to the nonperturbative structure of the QCD vacuum. The chiral soliton model, which is also called the linear sigma model (LSM) \cite{ref8}, has been proposed to describe the vacuum by incorporating chiral symmetry with its dynamical spontaneous breaking \cite{ref9,ref10,ref11,ref12,ref13} . This model could be solved in the mean-field approximation. There is a semiclassical solution which is referred to as a chiral soliton, and the hadron static properties can be easily derived from these solitons. \cite{ref14,ref15,ref16,ref17,ref18}.

In high energy physics, the thermal vacuum that is created by relativistic heavy ion collisions is very different from the vacuum at zero temperature and density \cite{ref19,ref20}. At high temperature or density the vacuum condensation melts, which results in chiral symmetry restoration and deconfinement phase transition of the system. Our purpose in this paper is to study the change of chiral soliton and its physical meaning in the whole chiral phase transition region. We introduce temperature and density within standard finite temperature field theory by embedding a soliton in a homogeneous hot and dense quark medium. The chiral soliton model provides a suitable working scheme to simultaneously study the restoration of chiral symmetry and the change of the soliton \cite{ref21}.

There is limited literature on the chiral soliton at finite temperature and density within the chiral soliton model \cite{ref21,ref22}. Abu-Shady and colleague~\cite{ref21} study solitons using an LSM at a finite temperature but not at a finite density in order to examine nucleon properties. Mao and colleagues~\cite{ref22} study the chiral solitons in the phase transition, but the authors only solve the chiral soliton solutions before the chiral restoration while neglecting the soliton solutions after the chiral restoration. They suggest that solitons do not exist after the phase transition.

It's generally held that the bound state of hadron melts away and the nucleon mass decreases to zero with chiral symmetry restoration; however, more recent research suggests that, while chiral condensation disappears after the chiral restoration, the nucleon mass does not go down and may increase, and the bound sate of quarks can survive even in the chiral symmetry restored phase~\cite{ref27,ref28,ref29}. We suggest that there are two different vacuums to account for in the chiral soliton model, the chiral symmetry broken vacuum in the chiral symmetry broken phase and the chiral restored vacuum in the chiral symmetry restored phase. In contrast to other research (for example~\cite{ref22}), we examine the chiral soliton solutions after the chiral restoration under the modified boundary conditions due to the chiral restored vacuum and analyse their physical properties. Therefore, the results presented in this paper are quite different from those obatained in reference \cite{ref22}, which will be discussed later.

The structure of this paper is as follows: in section II the chiral soliton model is introduced at zero as well as finite temperature and density. In section III, we present the thermal effective potential density, and show the soliton solutions of the chiral soliton equations at different temperatures and densities. The change of the soliton solutions in the chiral phase transition and its physical meaning are then discussed before the summary section.

\section{Model}
\subsection{The chiral soliton model at zero temperature and density}

The chiral effective Lagrangian of the $SU(2)_R \times SU(2)_L$ linear sigma model is \cite{ref14}
\begin{equation}
\mathcal L = \bar\psi [ i\gamma^\mu \partial_\mu + g( \hat\sigma + i \gamma_5 \vec\tau \cdot \vec{\hat\pi} )] \psi + \frac{1}{2}( \partial_\mu \hat\sigma \partial^\mu \hat\sigma + \partial_\mu \vec{\hat\pi} \partial^\mu \vec{\hat\pi}) - U( \hat\sigma, \vec{\hat\pi} ) ,\label{L}
\end{equation}
where $\psi$ represents the spin-$\frac{1}{2}$ two flavors light quark fields $\psi=(u,d)$, $\hat\sigma$ is the spin-0 isosinglet scalar field, and $\vec{\hat\pi}$ is the spin-0 isovector meson field $\vec{\hat\pi}=(\hat\pi_1,\hat\pi_2,\hat\pi_3 )$. The potential for $\hat\sigma$ and $\vec{\hat\pi}$ is
\begin{equation}
U ( \hat\sigma, \vec{\hat\pi} ) = \frac{\lambda}{4} (\hat\sigma^2 + \vec{\hat\pi}^2 - \nu^2 )^2 + H \hat\sigma - \frac{m_\pi^4}{4 \lambda} + f_\pi^2 m_\pi^2 ,\label{U}
\end{equation}
where the last two constant terms in equation (\ref{U}) are used to guarantee that the energy of a vacuum in the absence of quarks is zero. The minimum energy occurs for chiral fields $\hat\sigma$ and $\vec{\hat\pi}$ restricted to the chiral circle
\begin{equation}
\langle \hat\sigma \rangle^2 + \langle  \vec{\hat\pi}  \rangle^2 = f_\pi^2,
\end{equation}
where $f_\pi=93 MeV$ is the pion decay constant, $H \sigma$ is the explicit chiral symmetry breaking term, $H=f_\pi m_\pi^2$, and $m_\pi=138 MeV$ is the pion mass. Chiral symmetry is explicitly broken in the vacuum and the expectation values of the meson fields are: $\langle\sigma\rangle=-f_\pi$ and $\langle \vec{\hat\pi} \rangle=0$. The constituent quark mass in the vacuum is $M_q=g f_\pi$, and the $\sigma$ mass is defined by $m_\sigma^2=m_\pi^2+2\lambda f_\pi^2$. The quantity $\nu^2$ can be expressed as $\nu^2=f_\pi^2-m_\pi^2/\lambda$. In our calculation we follow Birse and colleagues~\cite{ref14} and set the constituent quark mass and the sigma mass as $M_q=500 MeV$ and $m_\sigma=1200 MeV$, that determine the parameters $g \approx 5.28$ and $\lambda \approx 82.1$.

From the Lagrangian~(\ref{L}) the field radial equations could be derived \cite{ref14,ref21}
\begin{eqnarray}
\frac{\mathrm{d} u(r)}{\mathrm{d} r} = -( \epsilon - g \sigma (r)) v(r) - g \pi(r) u(r) ,\label{u}\\
\frac{\mathrm{d} v(r)}{\mathrm{d} r} = -( \frac{2}{r} - g \pi(r)) v(r) + ( \epsilon + g \sigma(r)) u(r) ,\label{v}\\
\frac{\mathrm{d}^2 \sigma (r)}{\mathrm{d} r^2}  + \frac{2}{r} \frac{\mathrm{d} \sigma(r)}{\mathrm{d} r} + N g (u^2(r) - v^2(r)) = \frac{\partial U}{\partial \sigma} ,\label{sigma}\\
\frac{\mathrm{d}^2 \pi(r)}{\mathrm{d} r^2} + \frac{2}{r} \frac{\mathrm{d} \pi(r)}{\mathrm{d} r} - \frac{2\pi(r)}{r^2} + 2 N g u(r) v(r) = \frac{\partial U}{\partial \pi} .\label{pi}
\end{eqnarray}
In the above derivations one takes the mean-field approximation and the ``hedgehog" ansatz, which means
\begin{eqnarray}
&& \langle \hat\sigma (\vec{r}, t) \rangle = \sigma (r)  ,\ \ \ \ \langle \vec{\hat\pi} (\vec{r}, t) \rangle = \hat{\vec r} \pi(r) ,\\
&& \psi(\vec{r},t) = e^{-i \epsilon t} \sum_{i=1}^{N} q_i(\vec{r}),\ \ \ \ q(\vec{r}) = \binom{u(r)}{i \vec{\sigma} \cdot \hat{\vec{r}} v(r)} \chi ,\\
&& (\vec{\sigma} + \vec{\tau}) \chi = 0,
\end{eqnarray}
where $q_i$ are N identical valence quarks in the lowest s-wave level with eigen-energy $\epsilon$. $N$ is set to 3 for baryons and 2 for mesons. $\chi$ is the spinor. The quark functions should satisfy the normalization condition
\begin{equation}
4\pi \int r^2(u^2(r) + v^2(r)) \mathrm{d}r = 1 \label{n}.
\end{equation}
The boundary conditions are
\begin{eqnarray}
&& v(0)=0,\ \ \ \frac{\mathrm{d} \sigma(0)}{\mathrm{d} r}=0,\ \ \ \pi(0)=0, \label{b1}\\
&& u(\infty)=0,\ \ \ \sigma(\infty)=-f_\pi,\ \ \ \pi(\infty)=0 .\label{b2}
\end{eqnarray}
The asymptotic vacuum expectation value of the soliton field has to be determined by the condition that the physical vacuum is recovered at infinity.

The lowest quark energy eigenvalue is $\epsilon=30.5 MeV$. The equations (\ref{u}--\ref{pi}) together with normalization condition (\ref{n}) and boundary conditions (\ref{b1},\ref{b2}), which are nonlinear ordinary differential equations, could be numerically solved \cite{ref14}.

\subsection{The chiral soliton model at finite temperature and density}

In order to investigate the temperature and chemical potential dependence of the chiral soliton, we embed a single soliton in a homogeneous hot and dense quark medium with temperature $T$ and chemical potential $\mu$. First, we derive the effective potential of the spatially uniform system at finite temperature and density using the finite temperature field theory \cite{ref23}. The grand partition function is
\begin{eqnarray}
{\mathcal Z} &=& Tr \exp [-( \hat{\mathcal H} - \mu \hat{\mathcal N} )/T] \nonumber\\
&=& \int \prod_{j} {\mathcal D} \sigma {\mathcal D} \pi_j \int{\mathcal D} \psi {\mathcal D}\bar{\psi} \exp[\int_{x} ( {\mathcal L} + \mu \bar{\psi} \gamma^0 \psi )] ,
\end{eqnarray}
where $j=1,2,3$, $\int_x \equiv i \int_{0}^{1/T}{\mathrm{d} t} \int_V{\mathrm{d^3} x}$ and $V$ is the volume of the system.

In the mean-field approximation, we take the mesons fields $\sigma$ and $\pi$ as time-independent classical mean fields by their expectation values so that both the quantum and thermal fluctuations of the meson fields are neglected \cite{ref12}. The thermal effective potential of the homogeneous quark medium can be derived \cite{ref23}
\begin{equation}
\Omega (\sigma,\pi;T,\mu)= - \frac{T \ln \mathcal Z}{V} = U(\sigma,\pi ) + \Omega_{\bar{\psi} \psi} ,
\end{equation}
with the thermal quark and antiquark contribution
\begin{equation}
\Omega_{\bar{\psi} \psi} = -\nu_q T \int \frac{\mathrm{d^3} \vec{p}}{(2\pi )^3} \{ \ln[ 1+ e^{-(E_q - \mu)/T}] + \ln[ 1+ e^{-(E_q + \mu)/T}] \} ,
\end{equation}
where $\nu_q$ is the degeneracy factor $\nu_q=2(spin) \times 2(flavor) \times 3(color)=12$, $E_q=\sqrt{\vec p^2+M_q^2}$ is the valence quark and antiquark energy for $u$,$d$ quarks. The constituent quark (antiquark) mass $M_q$ is defined by
\begin{equation}
M_q^2 = g^2 (\sigma^2 + \pi^2 ) .
\end{equation}
The minimum energy either in a vacuum or in a thermal vacuum for chiral fields is restricted to the chiral circle
\begin{equation}
\sigma^2 + \pi^2 = \sigma_v^2 ,
\end{equation}
where the value of $\sigma_v$ in the thermal medium should be determined by the absolute minimum of the thermal effective potential, which is $\frac{\partial \Omega }{\partial \sigma}=0$ \cite{ref24}. Thus $\sigma_v$ changes with temperature or chemical potential.

Next we will embed one soliton into the homogeneous thermal background. For this purpose we first develop an effective Lagrangian,
\begin{equation}
 \mathcal L_{eff} = \bar\psi [ i \gamma^\mu \partial_\mu + g (\hat\sigma + i \gamma_5 \vec\tau \cdot \vec{\hat\pi} )] \psi +\frac{1}{2} ( \partial_\mu \hat\sigma \partial^\mu \hat\sigma +\partial_\mu \vec{\hat\pi} \partial^\mu \vec{\hat\pi} ) -\Omega (\sigma,\pi;T,\mu) ,\label{Leff}
\end{equation}
where the potential $U(\sigma,\pi)$ has been replaced by the thermal effective potential $\Omega$~\cite{ref21,ref22,ref30}. In a certain eigenstate of Hamiltonian \cite{ref14}, one could obtain its expectation value in the mean field approximation and the result is
\begin{equation}
\langle H \rangle = \int  \mathrm{d^3} \vec r \{ N q^{\dag} [-i \vec\alpha \cdot \vec\nabla -g\beta\sigma -i \gamma_5 \vec\tau \cdot \hat{\vec r}\pi] q + \frac{1}{2}|\vec\nabla \sigma|^2 + \frac{1}{2}|\vec\nabla \pi|^2 + \Omega (\sigma,\pi;T,\mu) \}.\label{H}
\end{equation}
where $\vec\alpha$ and $\beta$ are the Dirac matrices. $N$ is set to $3$ for baryon in the following discussion. Notice that the quark spinor wave function $q$ could be expressed by functions $u$ and $v$. The $u$, $v$, $\sigma$ and $\pi$ in this equation are all purely c-number quantities which could be determined by minimizing $\langle H \rangle$. Minimization of (\ref{H}) produces four coupled equations of quark and meson fields. The forms of the two equations about functions $u$ and $v$ are the same with the equations (\ref{u}) and (\ref{v}), while the other two equations about functions $\sigma$ and $\pi$ become
\begin{eqnarray}
\frac{\mathrm{d}^2 \sigma(r)}{\mathrm{d} r^2} + \frac{2}{r} \frac{\mathrm{d} \sigma(r)}{\mathrm{d} r} + N g (u^2(r) - v^2(r)) = \frac{\partial \Omega}{\partial \sigma} ,\\
\frac{\mathrm{d}^2 \pi(r)}{\mathrm{d} r^2} + \frac{2}{r} \frac{\mathrm{d} \pi (r)}{\mathrm{d} r} - \frac{2 \pi(r)}{r^2} + 2N g u(r) v(r) = \frac{\partial \Omega}{\partial \pi} ,
\end{eqnarray}
where
\begin{eqnarray}
\frac{\partial \Omega}{\partial \sigma} = \frac{\partial U(\sigma,\pi)}{\partial \sigma} + g^2 \sigma \nu_q \int \frac{\mathrm{d^3} p}{(2 \pi)^3} \frac{1}{E_q} (\frac{1}{1+ e^{(E_q - \mu)/T}} + \frac{1}{1+ e^{(E_q + \mu)/T}}) ,\label{sigma2}\\
\frac{\partial \Omega}{\partial \pi} = \frac{\partial U(\sigma,\pi)}{\partial \pi} + g^2 \pi \nu_q \int \frac{\mathrm{d^3} p}{(2 \pi)^3}\frac{1}{E_q} (\frac{1}{1+ e^{(E_q - \mu)/T}} + \frac{1}{1+ e^{(E_q + \mu )/T}}) .\label{pi2}
\end{eqnarray}
And the boundary conditions are
\begin{eqnarray}
&& v(0)=0 ,\ \ \ \frac{\mathrm{d} \sigma (0)}{\mathrm{d} r}=0 ,\ \ \ \pi(0)=0 , \label{b3}\\
&& u(\infty)=0 ,\ \ \ \sigma(\infty)=\sigma_v ,\ \ \ \pi (\infty )=0 . \label{b4}
\end{eqnarray}
At certain values of temperature and density, the soliton solutions can be obtained by numerically solving the quark field equations (\ref{u}, \ref{v}) and mesons field equations (\ref{sigma2}, \ref{pi2}) together with the normalization condition (\ref{n}) and the new boundary conditions (\ref{b3}, \ref{b4}).

The physical picture of our system should be reminded here. At zero temperature and density the soliton is embedded in a pure vacuum which is described by potential $U(\sigma, \pi)$. The physical vacuum is determined by $\frac{\partial U}{\partial \sigma}=\frac{\partial U}{\partial \pi}=0$. The $\sigma$ and $\pi$ fields assume their physical vacuum value at large radius $r$, which means inside the domain of the soliton the $\sigma$ and $\pi$ fields are $r$ dependent variables, while outside the domain of the soliton it is pure vacuum and the meson fields are constant background fields. At finite temperature and density the single soliton is embedded in a thermal vacuum which is described by thermal potential $\Omega(\sigma, \pi)$. The thermal physical vacuum is determined by $\frac{\partial \Omega}{\partial \sigma}=\frac{\partial \Omega}{\partial \pi}=0$. Inside the domain of the soliton the $\sigma$ and $\pi$ fields are $r$ dependent variables, while outside the domain of the soliton it is a thermal equilibrium background and the meson fields assume their physical thermal vacuum values.

The soliton solved out of the thermal background will minimize its energy (\ref{H}). Through the soliton solutions, the physical properties of the three-quark system can be calculated. The total energy or mass and the root mean square charge radius of the hedgehog baryon are respectively given by:
\begin{eqnarray}
E = M = N \epsilon + 4 \pi \int {\mathrm{d} r} r^2 [\frac{1}{2}(\frac{\mathrm{d} \sigma}{\mathrm{d} r})^2 + \frac{1}{2}(\frac{\mathrm{d} \pi}{\mathrm{d} r})^2 + \frac{\pi^2}{r^2} + \Omega(\sigma,\pi;T,\mu)] ,
\end{eqnarray}
\begin{eqnarray}
\langle R^2 \rangle= 4 \pi \int_0^\infty r^4 (u^2 + v^2) {\mathrm{d} r} .
\end{eqnarray}

It should be pointed out that the soliton is stable only when the total energy $E$ is lower than the energy of three free constituent quarks $3M_q$, because the equations always have the quasi-free particle plane wave solutions of quarks. This problem will be discussed later.

\section{phase transition and the soliton solutions}

\begin{figure}[tbh]
\includegraphics[width=210pt,height=150pt]{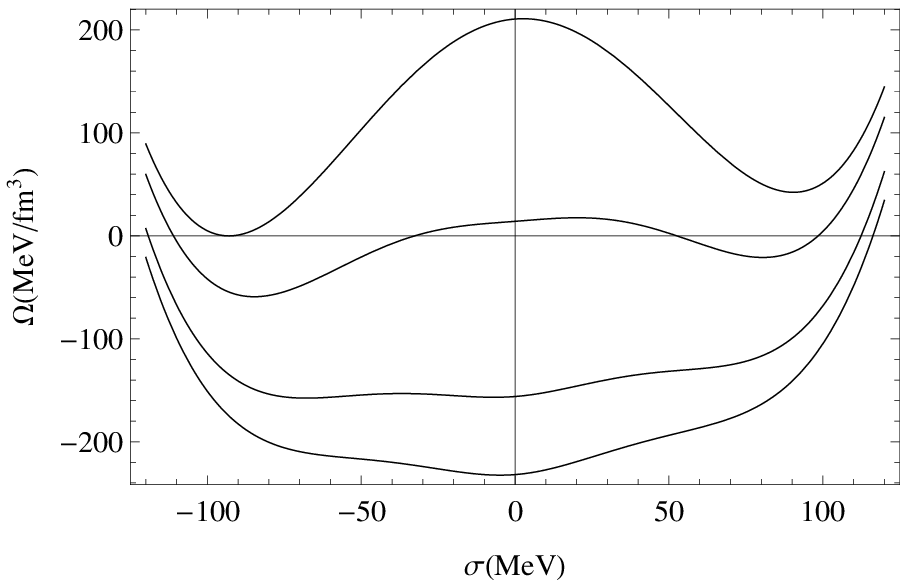}
\hspace{1cm}
\includegraphics[width=210pt,height=150pt]{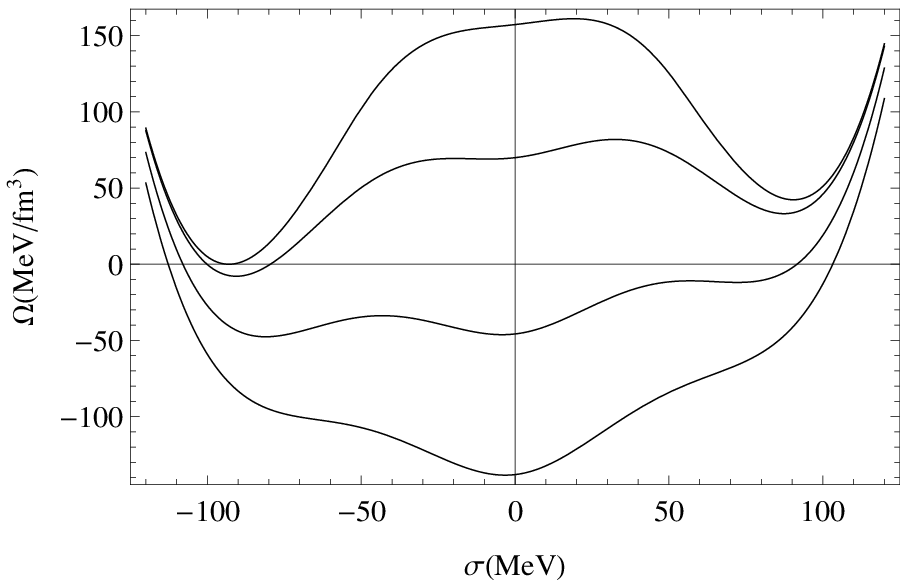}\\
(a)\; \; \; \; \; \; \; \; \; \; \; \; \; \; \; \; \; \; \; \; \;\; \; \; \; \; \; \; \; \; \; \; \; \; \; \; \; \;  (b)
\caption{The thermal effective potential $\Omega$, (a) at $\mu=0$, $T=\{0,160,187,196\} MeV$. (b) at $\mu=300 MeV$, $T=\{0,80,114,132\} MeV$.} \label{f1}
\end{figure}

It is instructive to plot the thermal effective potential as a function of the order parameter for different temperatures and densities. In Figure \ref{f1}, the left part (a) is for $\mu=0$ and the right (b) for $\mu=300 MeV$. One can clearly see how the chiral symmetry is restored. For $T=0$, there is only one absolute minimum of the potential and the chiral symmetry is entirely broken, so the absolute minimum corresponds to the chiral symmetry breaking (CSB) vacuum. As the temperature increases, another local minimum of the potential appears close to $\sigma=0$, which corresponds to the approximate chiral restored (ACR) vacuum. However, for $T<T_c$, the ACR vacuum is metastable, and the CSB vacuum is the physical vacuum. For $T=T_c$, the two vacuums are degenerate, so the chiral phase transition takes place at this time. As the temperature increases for $T>T_c$, the minimum of the ACR vacuum becomes the true minimum, thus the ACR vacuum is the physical vacuum, and the chiral symmetry is restored. It is clear that in this scenario the chiral phase transition is a first order phase transition.

\begin{figure}[tbh]
\includegraphics[width=210pt,height=150pt]{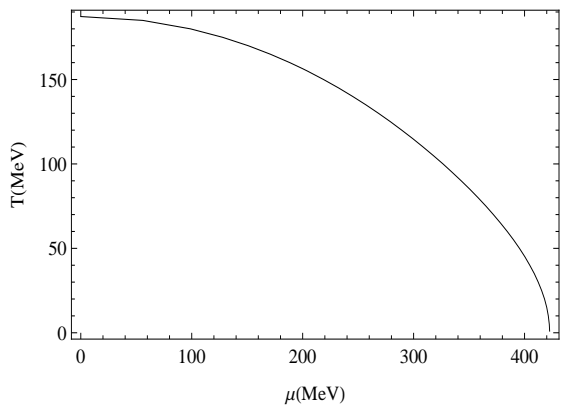}
\caption{The phase diagrams for the chiral soliton model in the $(T,\mu)$ plane.}\label{T-u}
\end{figure}

Phase diagrams for $(T,\mu)$ are presented in Fig.\ref{T-u}, which the points on the line corresponding to the states where the two vacuums are degenerate.

\begin{figure}[tbh]
\includegraphics[width=210pt,height=150pt]{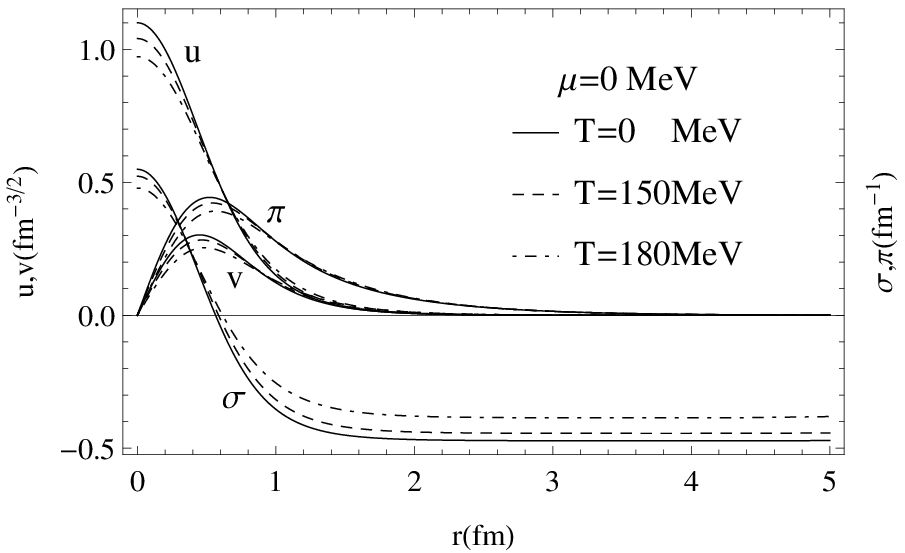}
\hspace{1cm}
\includegraphics[width=210pt,height=150pt]{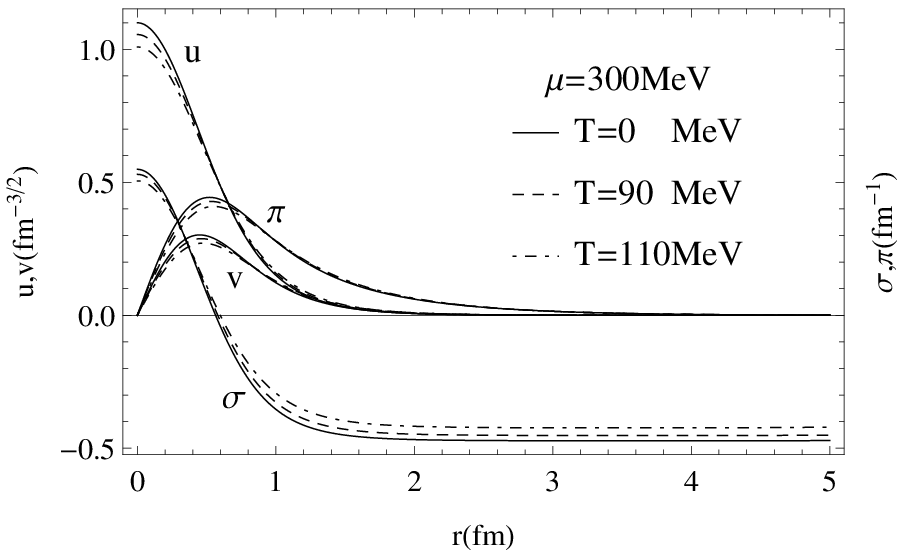}\\
(a)\; \; \; \; \; \; \; \; \; \; \; \; \; \; \; \; \; \; \; \; \;\; \; \; \; \; \; \; \; \; \; \; \; \; \; \; \; \; \; \; (b)
\caption{The quark fields $u(r)$, $v(r)$ and the meson fields $\sigma(r)$, $\pi(r)$ as functions of the radius $r$, (a) at $\mu=0 MeV$, $T=\{0,150,180\} MeV$. (b) at $\mu=300 MeV$, $T=\{0,80,110\} MeV$.}\label{f3}
\end{figure}

\begin{figure}[tbh]
\includegraphics[width=210pt,height=150pt]{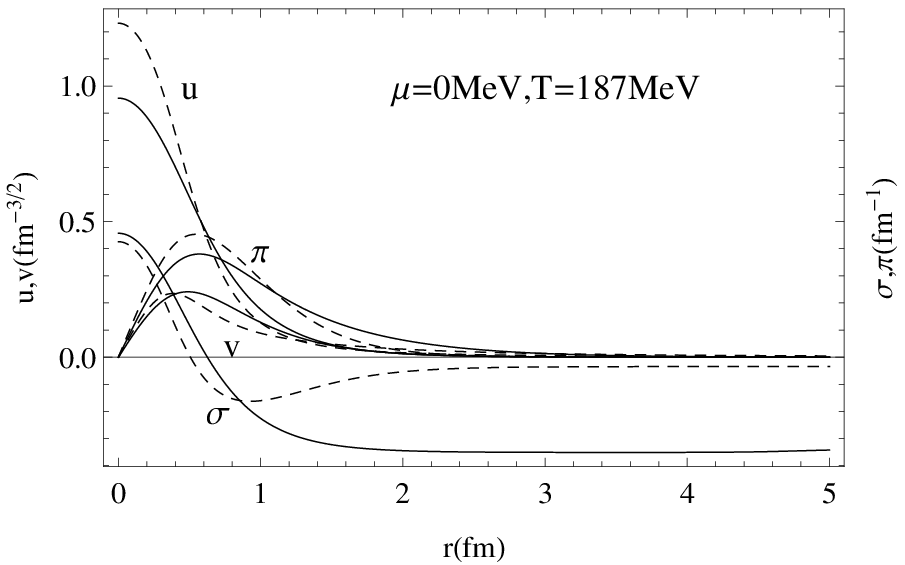}
\hspace{1cm}
\includegraphics[width=210pt,height=150pt]{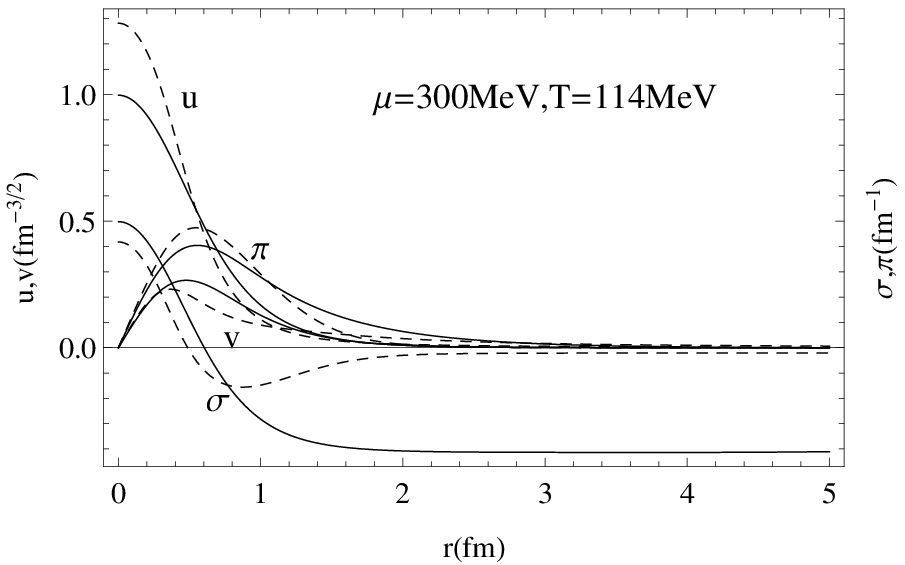}\\
(a)\; \; \; \; \; \; \; \; \; \; \; \; \; \; \; \; \; \; \; \; \;\; \; \; \; \; \; \; \; \; \; \; \; \; \; \; \; \; \; \; (b)
\caption{(a) The quark fields $u(r)$, $v(r)$ and the meson fields $\sigma(r)$, $\pi(r)$ as functions of the radius $r$ at $\mu=0$, $T=187 MeV$. (b) The quark fields $u(r)$, $v(r)$ and the meson fields $\sigma(r)$, $\pi(r)$ as functions of the radius $r$ at $\mu=300 MeV$, $T=114 MeV$. For both (a) and (b) the solid curves correspond to the boundary condition set for the CSB vacuum value, while the dashed curves correspond to the ACR vacuum value.} \label{f4}
\end{figure}

\begin{figure}[tbh]
\includegraphics[width=210pt,height=150pt]{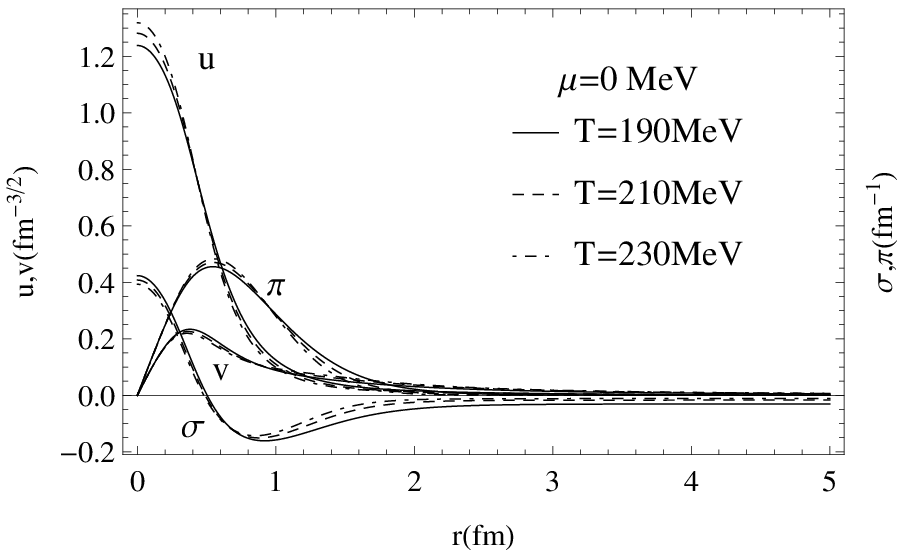}
\hspace{1cm}
\includegraphics[width=210pt,height=150pt]{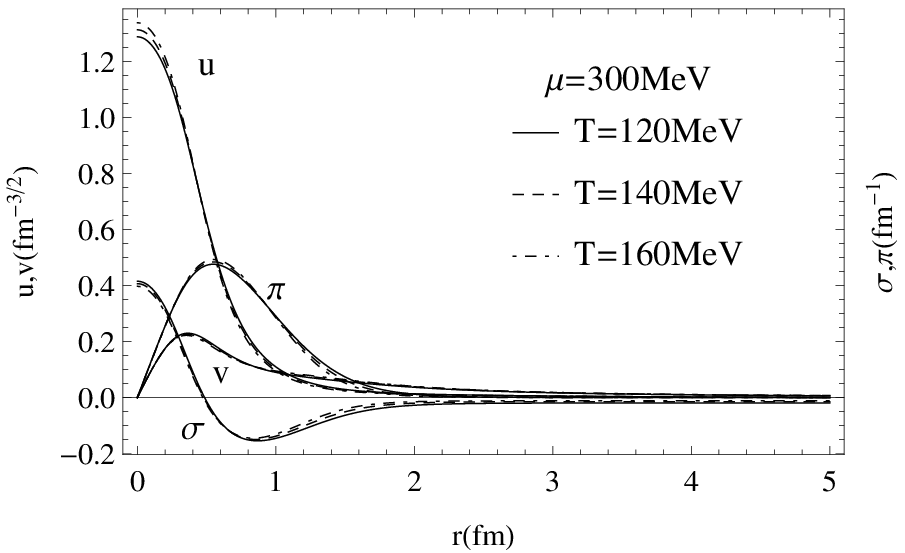}\\
(a)\; \; \; \; \; \; \; \; \; \; \; \; \; \; \; \; \; \; \; \; \;\; \; \; \; \; \; \; \; \; \; \; \; \; \; \; \; \; \; \; (b)
\caption{The quark fields $u(r)$, $v(r)$ and the meson fields $\sigma(r)$, $\pi(r)$ as functions of the radius $r$, (a) at $\mu=0$, $T=\{190,210,230\} MeV$. (b) at $\mu=300 MeV$, $T=\{120,140,160\} MeV$}. \label{f5}
\end{figure}

Next we study the soliton solutions through the chiral phase transition. We discuss two cases: one is holding the chemical potential fixed at zero and the other is holding the chemical potential fixed at $\mu=300 MeV$. In both cases, the temperature varies from low values to high values. The corresponding soliton solutions are represented in figure~\ref{f3}, figure~\ref{f4} and figure~\ref{f5}. The phase transition temperatures $T_c$ are $187 MeV$ and $114 MeV$ for $\mu=0$ and $\mu=300 MeV$, respectively.

For $T<T_c$, the CSB vacuum is the physical vacuum. In solving the field equations, one should notice that the $\sigma$ field approaches the value of the CSB vacuum as $r$ goes to infinity. In Fig.\ref{f3}, we plot the quark fields $u(r)$, $v(r)$ and the meson fields $\sigma(r)$, $\pi(r)$ as functions of the radius $r$ at fixed chemical potential for different temperatures: the left (a) is for $\mu=0$ and the right (b) for $\mu=300 MeV$. It can be seen that all amplitudes of the solitons decrease and change more and more rapidly as the temperature increases. By comparing the solitons at zero temperature between the two cases of $\mu=0$ and $\mu=300 MeV$, it can be seen that the shape of the soliton solutions do not change significantly, thus the amplitude of the soliton is not so sensitive to the variation of the chemical potential. These solutions represent quarks that are bounded in hadronic states in which the chiral symmetry is broken.

For $T=T_c$, the two vacuums are degenerate. The $\sigma$ field can approach either the CSB vacuum value or the ACR vacuum value as $r$ goes to infinity. Thus we solve the field equations using two different boundary conditions of the $\sigma$ field. In Fig.\ref{f4}, we plot the quark fields $u(r)$, $v(r)$ and the meson fields $\sigma(r)$, $\pi(r)$ as functions of the radius $r$ for the critical temperature and chemical potential: the left (a) is for $\mu=0$, $T=187 MeV$ and the right (b) for $\mu=300 MeV$, $T=114 MeV$. The solid lines represent the case that the $\sigma$ field at infinity approaches the value of the CSB vacuum, and the dashed line represent the case that the $\sigma$ field at infinity approaches the value of the ACR vacuum. Comparing the two kinds of solutions one can see that the $\sigma$ fields are substantially different as $r$ going to infinity due to the different vacuums. The amplitudes of the $\sigma$ fields in the case of the ACR vacuum are substantially smaller than those of the CSB vacuum, while the amplitudes of the $u$ and $\pi$ fields in the case of ACR vacuum are significantly larger. There is a chiral restoring phase transition at this time. A new bound state appears which is quite different from the hadronic state before the chiral restoration.

For $T>T_c$, the ACR vacuum is the physical vacuum. In solving the field equations, one observes that the soliton solution of the hadronic state disappears. In previous studies, this is regarded as a result of the delocalization of quarks. However, these studies have neglected the possible soliton solutions due to the new boundary conditions after the chiral restoration. As the physical vacuum is the ACR vacuum, the meson fields should assume the ACR vacuum values outside the domain of the soliton. Thus we solve the soliton equations at $T>T_c$ by choosing that the $\sigma$ field approaches the value of the ACR vacuum as $r$ goes to infinity. In Fig.\ref{f5}, we plot the quark fields $u(r)$, $v(r)$ and the meson fields $\sigma(r)$, $\pi(r)$ as functions of the radius $r$ at a fixed chemical potential for different temperatures, where the left (a) represents the case for $\mu=0$ and the right (b) for $\mu=300 MeV$. One can see that the amplitudes of the $\sigma$ fields decrease but the $u$ fields increase as the temperature increases, which is quite different from the case of $T<T_c$. Very different soliton solutions are present after the chiral restoration phase transition, which indicates a new bound state of quarks comes into being while the chiral symmetry is approximately restored. Shu and colleagues \cite{ref25} found a similar bound state of quarks after deconfinement phase transition in the Friedberg-Lee (FL) soliton model.

Next we will study the physical properties of these solitons by the above mentioned soliton solutions through the chiral phase transition. By subtracting the homogeneous medium contribution \cite{ref26}, the total effective energy $E^*$ of the soliton is given by the sum of the energy of the valence quarks and the kinetic energies of $\sigma$ field and $\pi$ field
\begin{equation}
E^* = N \epsilon +4\pi \int {\mathrm{d} r} r^2 [\frac{1}{2} (\frac{\mathrm{d} \sigma}{\mathrm{d} r})^2 + \frac{1}{2} (\frac{\mathrm{d} \pi}{\mathrm{d} r})^2 + \frac{\pi^2}{r^2}].
\end{equation}

\begin{figure}[tbh]
\includegraphics[width=210pt,height=150pt]{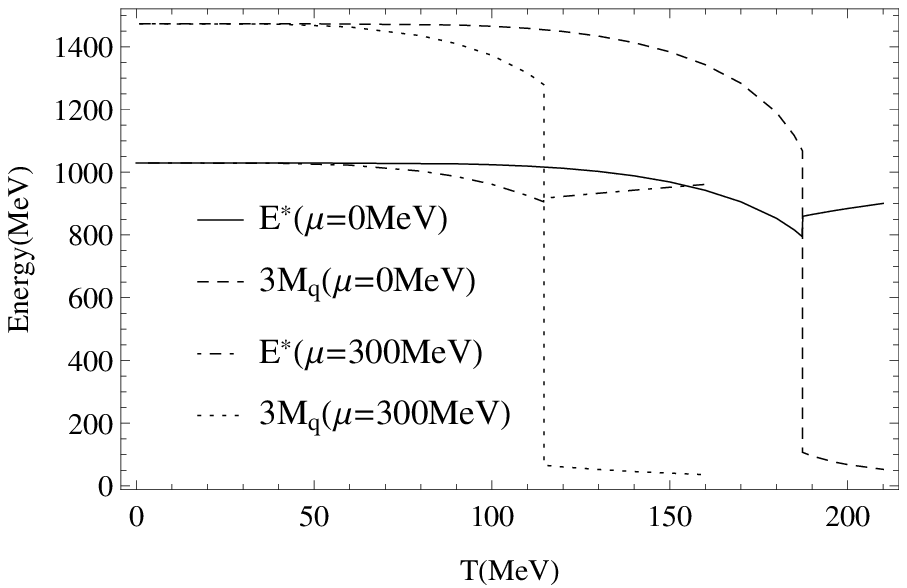}
\hspace{1cm}
\includegraphics[width=210pt,height=150pt]{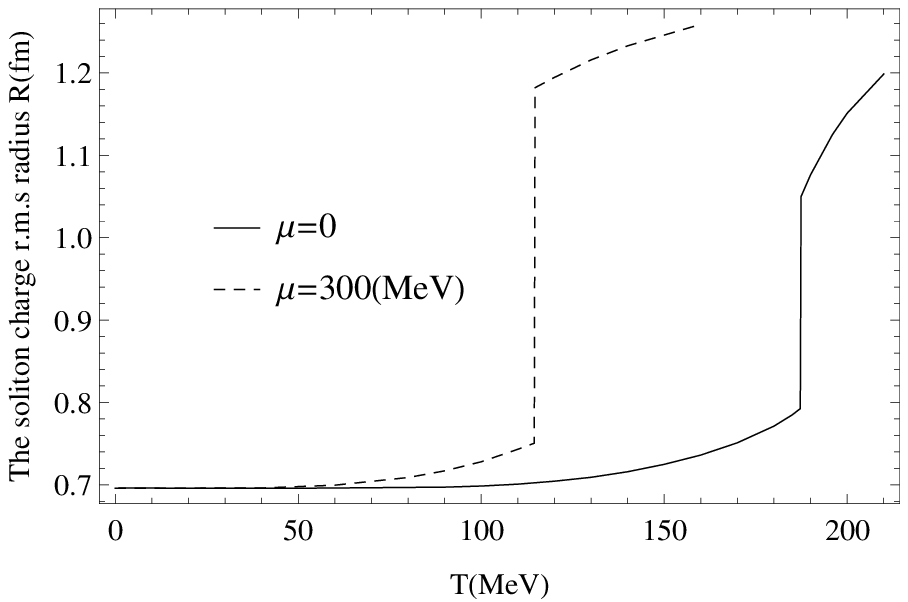}\\
(a)\; \; \; \; \; \; \; \; \; \; \; \; \; \; \; \; \; \; \; \; \; \; \; \; \; \; \; \; \; \; \; \; \; \; \; \; \; \; \; (b)
\caption{ (a) The total effective energy of a stable chiral soliton $E^*$ and the energy of three free constituent quarks mass $3M_q$ as a function of the temperature $T$ at $\mu=0$ and $\mu=300 MeV$. (b) The charge r.m.s radius of a stable chiral soliton as a function of temperature $T$ at $\mu=0$ and $\mu=300 MeV$.} \label{f6}
\end{figure}

In Fig.\ref{f6}(a), the total effective energy $E^*$ of the soliton and the energy of three free constituent quarks $3M_q$ are plotted as functions of the temperature $T$ for $\mu=0$ and $\mu=300 MeV$. One can see that the soliton energy $E^*$ decreases more and more rapidly with the temperature increases for $T<T_c$. The energy of three free constituent quarks mass $3M_q$ has the similar behavior, and it is larger than $E^*$, which means that the soliton solutions have the lowest energy and the soliton is stable. As the temperature approaches the chiral phase transition temperature $T_c$, $E^*$ suddenly jumps to a high value. The sudden leap reflects the first-order chiral phase transition and also signals a drastic structural change in the quark system. At $T_c$, $3M_q$ suddenly jumps to an extremely low value near zero. When $T>T_c$, the chiral symmetry is approximately restored, $E^*$ increases slowly with the temperature increasing, while the $3M_q$ keeps in the extremely low value.

One can see that after chiral restoration the $3M_q$ is lower than $E^*$, which means that the soliton solutions are only the local minimum of the energy. However we can not simply throw these soliton solutions away because the quark masses decreasing near $T_c$ is controversial. Brown-Rho scaling suggests that quark masses and such hadrons as $\rho$, $N$ decrease when chiral symmetry gets restored. Yet a number of empirical facts contradicts to this idea. Shuryak and colleagues argue, in recent studies \cite{ref27,ref28}, that several observations indicate that the nucleon mass does not go down and even seems to be increasing as $T$ grows through $T_c$. The authors point out that by adding a chirally-even mass  $m_\chi\sim \bar q\partial_0 \gamma_0 q$ to the constituent quark mass the total quark mass appears to be less affected by chiral restoration.  From Fig.\ref{f6}(a), one can see that our result is consistent with this because the soliton mass through the chiral phase transition slowly decreases and then slowly increases with temperature increasing through $T_c$. And the recent lattice studies \cite{ref29} also report the similar results. Our results also suggest that the chirally-even mass should be considered in studying the chiral restoration in this model, which is an interesting topic worthy of further study.

In Fig.\ref{f6}(b) we illustrate the charge root mean square (r.m.s) radius $R$ of the chiral soliton as a function of the temperature $T$ for $\mu=0$ and $\mu=300 MeV$. In both cases, the soliton radius $R$ increases with the rise in temperature, and it shows a swelling of the soliton when temperature and density increase. As the temperature is close to the chiral phase transition temperature $T_c$, $R$ suddenly jumps to a high value, which also indicates a drastic structural change in the system. After the $T_c$, $R$ increases more quickly than before the $T_c$ with the increase in temperature, which also suggests that the properties of this bound state of quarks after the chiral restoration are quite different from those of the hadronic bound state.

\section{Summary and discussion}

In this paper, we have studied the chiral soliton model at finite temperature and density, and solved the chiral soliton equations at different temperatures and densities with proper boundary conditions. Different soliton solutions are discussed through the chiral phase transition. Our results show that there are also soliton solutions after the chiral restoration. Our results, which differ from those derived by Mao and colleagues \cite{ref22}, show that there are soliton solutions after the chiral restoration. Mao and colleagues assert \cite{ref22} that solitons are melted away after the chiral phase transition, which takes a signal for the delocalization of the quarks. Our findings before the chiral restoration are  consistent with Mao and colleagues \cite{ref22}, but after the chiral restoration, because of the boundary conditions that we set, the soliton solutions do not disappear. As the vacuum changed due to the chiral symmetry restoration, we take the physical vacuum of the chiral restored phase as the boundary conditions of the meson fields. The soliton equations are solved under the modified boundary conditions, and the soliton solutions do exist after the chiral restoration. The results suggest that the quarks are still bounded even after the chiral restoration. The physical properties of these bound states are very different from those of the hadronic bound states. We argue that the new type of soliton solutions after the chiral restoration is important in understanding the slow increase of nucleon mass after the chiral restoration of recent lattice results. However, more physical properties of these new solitons deserve further study.

\begin{acknowledgments}
This work was supported in part by the National Natural Science Foundation of China with No. 10905018 and No. 11275082.
\end{acknowledgments}


\begin{thebibliography}{}

\bibitem{ref1}
D.H.~Rischke, Prog. Part. Nucl. Phys. 52 (2004) 197.
\bibitem{ref2}
K.~Yagi, T.~Hatsuda and Y.~Miake, ``Quark-gluon plasma: From big bang to little bang," Camb.Monogr. Part. Phys. Nucl. Phys. Cosmol. 23 (2005) 1.
\bibitem{ref3}
S.J.~Dong, J.-F.~Lagae, K.F.~Liu, Phys. Rev. Lett. 75 (1995) 2096.
\bibitem{ref4}
T.D.~Cohen, Prog.~Part. Nucl. Phys. 35 (1995) 221.
\bibitem{ref5}
R.~Alkofer, L.~Smekal, Phys. Rep. 353 (2001) 281.
\bibitem{ref6}
U.~Meissner, Rep.~Prog. Phys. 56 (1993) 903.
\bibitem{ref7}
M.S.~Bhagwat, L.~Chang, Y.X.~Liu, C.D.~Roberts, P.C.~Tandy, Phys. Rev. C76 (2007) 045203.
\bibitem{ref8}
M.~Gell-Mann and MLevy, Nuovo Cim. 16 (1960) 705.
\bibitem{ref9}
G.~Amelino-Camelia and S.Y.~Pi, Phys. Rev. D47 (1993) 2356.
\bibitem{ref10}
N.~Pertropoulos, J. Phys. G25 (1999) 2225.
\bibitem{ref11}
J.T.~Lenaghan, D.H.~Rischke and J.~Schaffner-Bielich, Phys. Rev. D62 (2000) 085008.
\bibitem{ref12}
O.~Scavenius, A.~Mocsy, I.N.~Mishustin and D.H.~Rischke, Phys. Rev. C64 (2001) 045202.
\bibitem{ref13}
S.~Shu and J.-R.~Li, J. Phys. G31 (2005) 459.
\bibitem{ref14}
M.C.~Birse and M.K.~Banerjee, Phys. Lett. B136 (1984) 284; M.C.~Birse and M.K.~Banerjee, Phys. Rev. D31 (1985) 118.
\bibitem{ref15}
T.D.~Cohen and W.~Broniowski, Phys. Rev. D34 (1986) 3472.
\bibitem{ref16}
K.~Goeke, M.~Harvey, F.~Gr¡§ummer and J.N.~Urbano, Phys. Rev. D37 (1988) 754.
\bibitem{ref17}
T.S.T.~Aly, J.A.~McNeil and S.~Pruess, Phys. Rev. D60 (1999) 114022.
\bibitem{ref18}
W.~Broniowski and B.~Golli, Nucl. Phys. A714 (2003) 575.
\bibitem{ref19}
M.~Gyulassy and L.~McLerran, Nucl. Phys. A750, (2005) 30.
\bibitem{ref20}
J.I.~Kapusta, J. Phys. G34 (2007) S295-304.
\bibitem{ref21}
M.~Abu-Shady and H.M.~Mansour, Phys. Rev. C85 (2012) 055204.
\bibitem{ref22}
H.~Mao, T.-Z.~Wei and J.-S.~Jin, Phys.Rev. C88 (2013) 035201.
\bibitem{ref27}
Shuryak E. On chiral symmetry breaking, topology and confinement[J]. Nuclear Physics A, 2014.
\bibitem{ref28}
J. Liao and E. V. Shuryak, Phys. Rev. D 73 (2006) 014509
\bibitem{ref29}
L. Y. Glozman, C. B. Lang and M. Schrock, Phys. Rev. D 86 (2012) 014507
\bibitem{ref23}
J.I.~Kapusta and C.~Gale, ``Finite-Temperature Field Theory: Principle and Aplications" (Cambridge University Press, UK, 2006).
\bibitem{ref24}
L.P.~Csernai and I.N.~Mishustin, Phys. Rev. Lett 74 (1995) 5005.
\bibitem{ref30}
S.~Shu and J.-R.~Li, Int. J. Mod. Phys. Conf. Ser. 29 (2014) 1460213
\bibitem{ref25}
S.~Shu and J.-R.~Li, Phys. Rev. C82 (2010) 045203.
\bibitem{ref26}
J.~Berger and C.-V.~Christov, Nucl. Phys. A 609 (1996) 537.

\end{thebibliography}
\end{document}